\documentclass[12pt,epsf,amstex]{article}
\usepackage [dvips]{graphicx}
\usepackage{amsmath}
\usepackage{amssymb}
\usepackage{psfig}
\usepackage{epsfig}

\addtocounter{secnumdepth}{1}
\begin{document}
\newcommand{\br}{\bar{r}}
\newcommand{\bbeta}{\bar{\beta}}
\newcommand{\bgamma}{\bar{\gamma}}
\newcommand{\bR}{{\bf{R}}}
\newcommand{\bS}{{\bf{S}}}
\newcommand{\half}{\frac{1}{2}}
\newcommand{\summ}{\sum_{m=1}^n}
\newcommand{\sumqno}{\sum_{q\neq 0}}
\newcommand{\tsum}{\Sigma}
\newcommand{\bsA}{\mathbf{A}}
\newcommand{\bsV}{\mathbf{V}}
\newcommand{\bsE}{\mathbf{E}}
\newcommand{\bsZ}{\hat{\mathbf{Z}}}
\newcommand{\bse}{\mbox{\bf{1}}}
\newcommand{\bspsi}{\hat{\boldsymbol{\psi}}}
\newcommand{\cdottt}{\!\cdot\!}
\newcommand{\deltaR}{\delta\mspace{-1.5mu}R}

\newcommand{\bGamma}{\boldmath$\Gamma$\unboldmath}
\newcommand{\dd}{\mbox{d}}
\newcommand{\ee}{\mbox{e}}
\newcommand{\p}{\partial}

\newcommand{\la}{\langle}
\newcommand{\ra}{\rangle}
\newcommand{\rao}{\rangle\raisebox{-.5ex}{$\!{}_0$}}  
\newcommand{\rae}{\rangle\raisebox{-.5ex}{$\!{}_1$}}
\newcommand{\raG}{\rangle_{_{\!G}}}
\newcommand{\rainr}{\rangle_r^{\rm in}}
\newcommand{\beq}{\begin{equation}}
\newcommand{\eeq}{\end{equation}}
\newcommand{\bea}{\begin{eqnarray}}
\newcommand{\eea}{\end{eqnarray}}
\def\lsim{\:\raisebox{-0.5ex}{$\stackrel{\textstyle<}{\sim}$}\:}
\def\gsim{\:\raisebox{-0.5ex}{$\stackrel{\textstyle>}{\sim}$}\:}

\numberwithin{equation}{section}

\thispagestyle{empty}
\title{\Large {\bf Planar Voronoi cells:\\[2mm]
the violation of Aboav's law explained}}

\author{{\bf H.\,J. Hilhorst}\\[5mm]
{\small Laboratoire de Physique Th\'eorique,
B\^atiment 210, Universit\'e de Paris-Sud}\\[-1mm]
{\small 91405 Orsay Cedex, France}\\}

\maketitle
\begin{small}

\begin{abstract}
In planar cellular systems $m_n$ denotes the
average sidedness of a cell neighboring an $n$-sided cell.
Aboav's empirical law states that $nm_n$ is linear in $n$.
A downward curvature is nevertheless
observed in the numerical $nm_n$  data of the Random Voronoi Froth.
The exact large-$n$ expansion of $m_n$ obtained in the present work,
{\it viz.} $m_n=4+3(\pi/n)^{\frac{1}{2}}+\ldots$, explains this curvature.
Its inverse square root dependence on $n$ sets a new theoretical
paradigm. 
Similar curved behavior may be expected, and must indeed be looked for,
in experimental data of sufficiently high resolution.
We argue that it occurs, in particular,
in diffusion-limited colloidal aggregation
on the basis of recent simulation data
due to Fern\'andez-Toledano {\it et al.} 
[{\it Phys.~Rev.~E\,} {\bf 71}, 041401 (2005)] and earlier
experimental results by Earnshaw and Robinson
[{\it Phys.~Rev.~Lett.} {\bf 72}, 3682 (1994)].
\end{abstract}

\end{small}
\vspace{60mm}

\noindent LPT -- ORSAY 05/89\\
{\small $^1$Laboratoire associ\'e au Centre National de la
Recherche Scientifique - UMR 8627}
\newpage


\section{Introduction}


\subsection{General}
\label{secgeneral}

In nature, planar cellular systems come in a wide
variety. They include biological tissues \cite{Lewis,MAI93,JeuneBarabe98},
polycrystals \cite{Aboav70},
cells formed by particles trapped at a water/air interface
\cite{MejiaRosalesetal00}, cells in surface-tension driven B\'enard convection 
\cite{CRR96}, in two-dimensional soap froths \cite{GGS87},
and in magnetic liquid froths \cite{Eliasetal97}.
In other systems cellular structure may appear when the data are subjected
to the Voronoi construction \cite{Okabeetal00}.
Examples of these are hard disks on an air table \cite{Lemaitreetal93},
a binary liquid during late stage coarsening \cite{SMS94},
two-dimensional colloidal aggregation
\cite{EarnshawRobinson94,EarnshawRobinson95,EHR96},
nanostructured cellular layers \cite{MTB02},
and studies of two-dimensional melting \cite{ZLM99}.

Two empirical rules play a prominent
role in studies of planar cellular systems:
Lewis' law and Aboav's law.
Both are statements about the statistics of a cell's
most conspicuous properties, {\it viz.} its area and its number of sides.
Lewis' law \cite{Lewis} 
says that the average area $A_n$ of an $n$-sided cell 
increases with $n$ as 
\beq
A_n=\frac{a_0}{\lambda}(n-n_0),
\label{eqnAn}
\eeq
where $a_0$ and $n_0$ are constants and
$\lambda$ is the areal cell density. 
In the present work 
we are interested in the second one of these laws, 
formulated by Aboav \cite{Aboav70}, 
who noticed that many-sided cells tend to have few-sided
neighbors and {\it vice versa.} He expressed this correlation in terms of 
the average $m_n$ of the
number of sides of a cell that neighbors an
$n$-sided cell. Aboav's law, 
also called the Aboav-Weaire \cite{Weaire74} law, asserts that
\beq
m_n=6-a+\frac{b}{n}\,,
\label{eqnmn}
\eeq
where $a$ and $b$ are numerical constants. This 
law, which expresses an intuitively plausible trend with $n$,
is in widespread use
\cite{MAI93,JeuneBarabe98,MejiaRosalesetal00,CRR96,GGS87,Eliasetal97,
Lemaitreetal93,SMS94,EarnshawRobinson94,MTB02,ZLM99}  
in the analysis of experimental data on cellular structures.
One usually plots $\,nm_n=(6-a)n+b\,$ as a function of $n$ 
and often refers to either this relationship or 
equation (\ref{eqnmn}) as the {\it linear law}.  

The surprising fact is that for many of these cellular systems,
in spite of all their diversity, Aboav's linear law 
appears to hold with good accuracy in 
most of the experimentally accessible range, 
which runs from $n=3$ to $n$ typically between $9$ and $12$. 
The experimental values of the Aboav parameters,
listed {\it e.g.} in reference \cite{ZsoldosSzasz99},
are typically in the range $0.7 \lesssim a \lesssim 1.5$
and $5.7 \lesssim b \lesssim 8.5$ \cite{footnote3}.
Attempts to explain the linearity with $n$ and the 
numerical values of $a$ and $b$ from first principles, {\it i.e.,} 
on the basis of a microscopic geometrical model, have been unsuccessful.
The various derivations of (\ref{eqnmn}) 
that one does find in the literature 
all involve approximations
(usually of the mean-field kind; see {\it e.g.} 
reference \cite{EdwardsPithia94}) 
or hypotheses whose general validity is subject to caution 
(we consider the `maximum entropy method' 
\cite{RivierLissowski82} to be in this class). 
The dominant view today is probably that (\ref{eqnmn}) is not exact,
but merely a good approximation to some unknown `true' curve,
which need not be the same for all cellular systems.

It has been realized, 
in particular by Le Ca\"er and Delannay \cite{LeCaerDelannay93}, 
that {\it knowledge about the large-$n$ behavior of} $m_n$ 
will constrain the law for $m_n$,
for example 
by establishing the regime of validity of its linearity
or by putting limits on the numerical values of $a$ and $b$.
The question of the large-$n$ behavior of $m_n$ is however a difficult one
and has received little attention since.
The `derivations' referred to above generally lead to (\ref{eqnmn})
without providing any restrictions on its range of validity.
In the present work we return to the large-$n$ behavior. We analyze
it within the context of the Random Voronoi Froth and
consider the implications of our conclusion for the interpretation of
simulations and experiments.
\vspace{2mm}


\subsection
{The Random Voronoi Froth (RVF)}
\label{secRVF}

We consider the `Random Voronoi Froth' (RVF),
in the mathematical literature rather called
the `Poisson-Voronoi tessellation'.
It is obtained by constructing the Voronoi cells
\cite{Okabeetal00} of a planar set of
randomly and uniformly distributed point centers,
for convenience often called `seeds' (but without
the implication that they are material). 
The statistical properties  of the RVF
have been studied 
more than those of any other microscopic model of cellular structure.
Reference \cite{Okabeetal00} compiles
a large body of analytic results and numerical tables. 
For the role and place of the RVF amidst other models of cellular systems 
one may consult, {\it e.g.,} Rivier \cite{Rivier93} or
Schliecker \cite{Schliecker02}.

Analytically, a first-principle derivation of 
an exact expression for $m_n$ in the RVF
(or, for that matter, in any other geometrical model)
is still lacking. 
The reason is that the calculation of this quantity
is in the notoriously difficult class of many-body problems.
Numerically, however,  
the $m_n$ values of the RVF are known with considerable 
precision \cite{BootsMurdoch83,LeCaerHo90,KumarKurtz93,Brakke}.
The best data come from
Monte Carlo simulations by Brakke \cite{Brakke}, whose results
for $m_n$ have a four digit accuracy in the range $4 \leq n \leq 9$.
In figure \ref{Fig1} we show these
data as well as those by
Boots and Murdoch \cite{BootsMurdoch83}, represented in the usual way
in an $nm_n$ {\it versus} $n$ plot.
The dashed straight line is Aboav's law 
(\ref{eqnmn}) with $a=0.75$ and $b=5.76$.
Although this law provides what appears as a very good approximation, 
the data points nevertheless exhibit
an extremely small but distinctive downward curvature
and the narrowness of their error bars rules out Aboav's linear fit.
We have computed the second derivative
$(nm_n)''\equiv(n-1)m_{n-1}-2nm_n+(n+1)m_{n+1}$ from Brakke's \cite{Brakke}
numerical data and plotted it in figure \ref{Fig1}.
The ratio $-(nm_n)''/(nm_n)$, 
which may be taken as a measure of the curvature, is 
of the order of only $1/250=0.004$.  
Hence a successful theory of the RVF should explain 
the existence as well as the sign and the smallness of the curvature.
So far, in order to accommodate this departure from linearity, only
{\it ad hoc\,} alternatives
to (\ref{eqnmn}) have been proposed
(see, {\it e.g.}, the discussions in \cite{LeCaerHo90} and in
section 5.3.3 of \cite{Okabeetal00}).

We pursue here the approach initiated in
reference \cite{HJHletter05}, where 
the methods of statistical mechanics were brought 
to bear on planar Voronoi tessellations. 
That work has opened the possibility of an 
expansion in powers of $n^{-\frac{1}{2}}$
for all quantities of interest related to an $n$-sided Voronoi cell.
This initially led to the asymptotic large-$n$ expansion of the
probability $p_n$ for an arbitrarily chosen cell to have $n$ sides. 
An immediate further result was the proof that the RVF
obeys Lewis' law (\ref{eqnAn})
{\it asymptotically for} $n\to\infty$ with coefficient 
$a_0=\frac{1}{4}$.
But whereas Lewis' law refers to a single cell, Aboav's deals with the
intrinsically more difficult problem of cell-cell correlations;
it therefore requires the separate study that we present here.

Our key result, equation (\ref{mnaspt}) of section \ref{secfirstneighbor},
is an exact expression for the asymptotic behavior of $m_n$ as
$n$ becomes large. The inverse square root decay appearing in that expression
explains the small deviations from linearity that are observed.
As a corollary we obtain in section \ref{secsecondneighbor} 
the asymptotic expression
for the total number, to be called $K_n^{(2)}$, 
of second-neighbor cells surrounding a central
$n$-sided cell. 
In section \ref{secdiscussion} we discuss these results.
We compare them to the numerical RVF data,
as well as to data on diffusion limited colloidal aggregation coming from
both experiments and simulations.
Whereas our asymptotic expansion constitutes an exact result based on first
principles, in section \ref{secnumerical} we take a pragmatic attitude and
try to find the best two- and three-parameter fits for $m_n$
that incorporate this large $n$ behavior. 
In section \ref{secnonRVF} we speculate briefly on
modifications that our theory may undergo in the case of 
cellular systems not of the RVF type.
In section \ref{secconclusion} we conclude.
Finally, in appendix A we prove a theorem concerning 
the statistics of randomly distributed points in a half-plane
that is indispensible for the derivation of the main result.


\section{Two-cell correlations}
\label{sectwocellcorrelations}


\subsection{First-neighbor correlation}
\label{secfirstneighbor}

We will determine analytically
the large-$n$ behavior of $m_n$ in a RVF.
Two steps are required to arrive at this result. The first step of the
derivation is based on the geometrical properties of the large
$n$-sided cell; these have all been determined quantitatively
in earlier work \cite{HJHletter05}, where they required 
considerable mathematical effort. 
The arguments of the first step will be
presented with the aid of a figure, but we stress that they are
nonetheless nontrivial and dictated by compelling logic; 
they constitute the conceptually new part of this investigation. 
The second step is a problem in statistics whose solution, 
although somewhat lengthy, can be obtained by standard methods.
We present it in appendix \ref{secappendix}. 
\begin{figure}
\vspace{5mm}
\scalebox{.45}
{\includegraphics{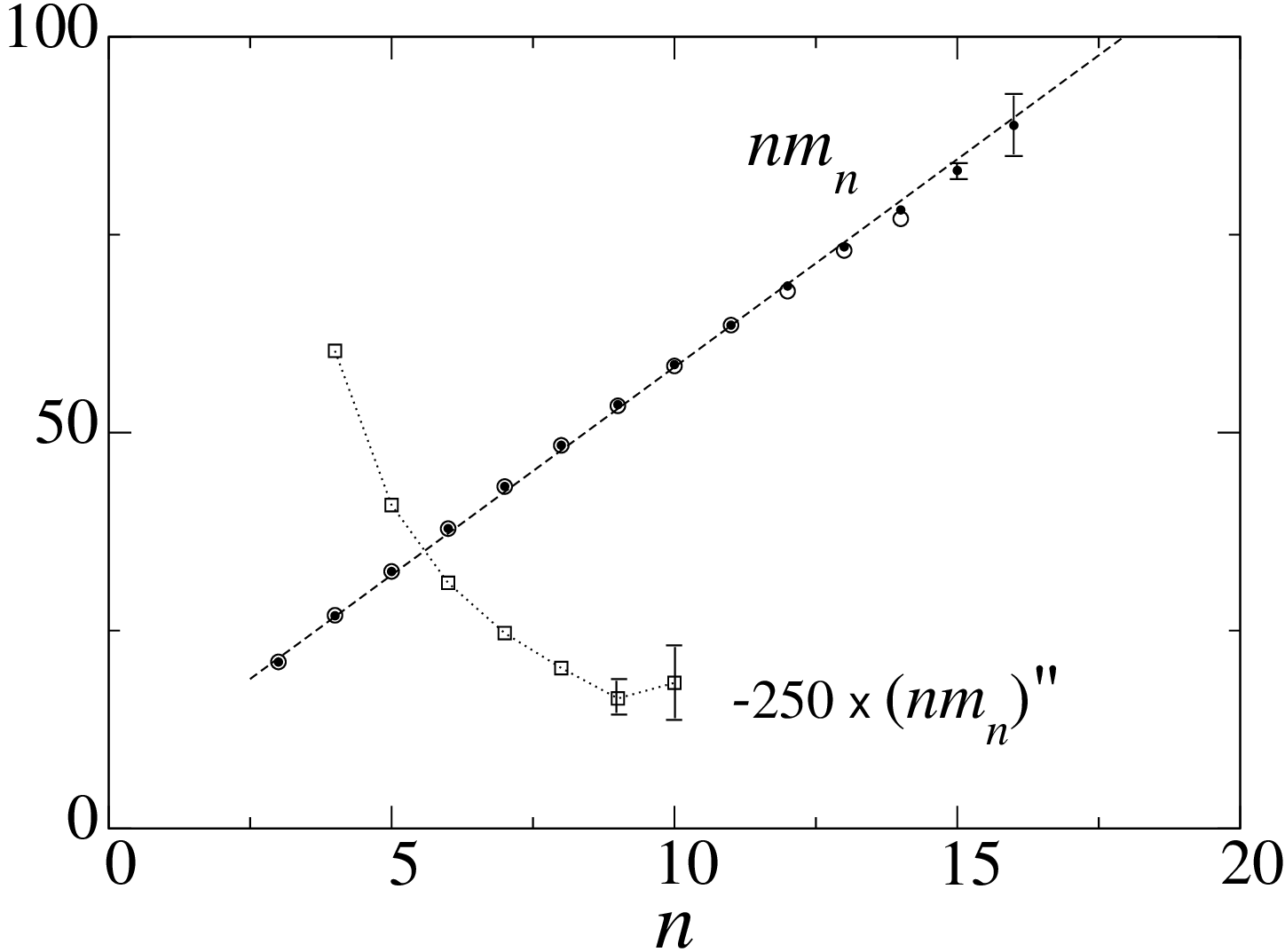}}
\caption{\small The cells neighboring an $n$-sided cell have together 
a total average of $nm_n$ sides. 
Open circles: simulation values of $nm_n$ according to Boots and
Murdoch \cite{BootsMurdoch83}. Dots: simulation values 
of $nm_n$ according to Brakke \cite{Brakke}. 
Dashed line: Aboav's linear law $nm_n=(6-a)n+b$ 
with $a=0.75$ and
$b=5.76$. Open squares: the second derivative $(nm_n)''\equiv
(n-1)m_{n-1}-2nm_n+(n+1)m_{n+1}$, constructed from Brakke's \cite{Brakke}
$m_n$ data and multiplied by $-250$ in order to be visible on
the scale of the figure. 
Error bars are shown for the dot and open square data; where absent, they
are smaller than the data point symbols.}
\label{Fig1}
\end{figure}
\begin{figure}
\vspace{5mm}
\scalebox{.45}
{\includegraphics{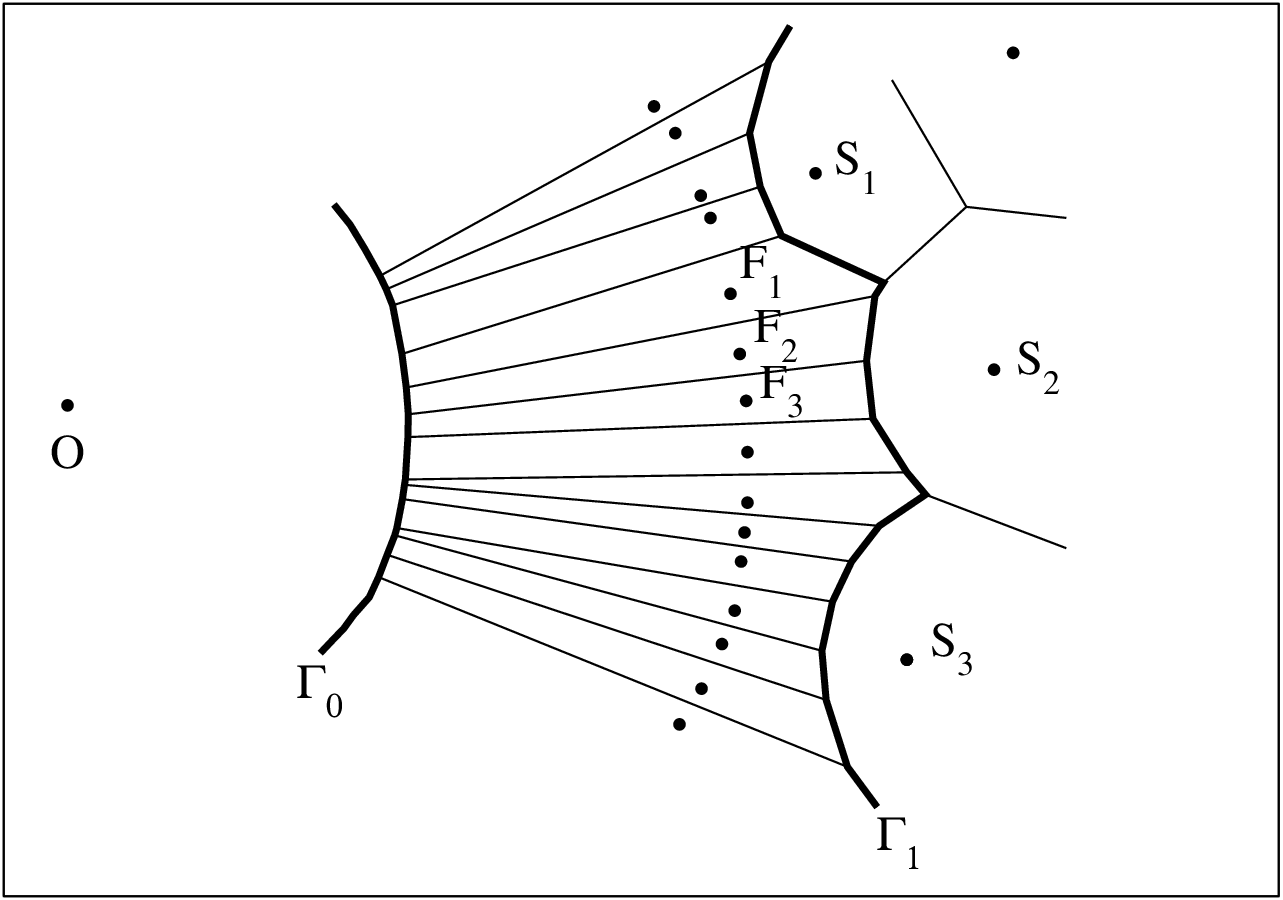}}
\caption{{\small 
Schematic picture of the environment of an $n$-sided Voronoi cell 
with $n \approx 100$ around a seed at $O$. 
The first and second neighbor cells 
have their seeds at $F_1, F_2, F_3, \ldots$ and 
at $S_1,S_2,S_3,\ldots$, respectively.
All solid line segments separate Voronoi cells. Among these,
the heavy solid line $\Gamma_0$ is the perimeter of the $n$-sided cell,
which is close to circular.
The heavy solid line $\Gamma_1$  separates the first from the second
neighbors. Both $\Gamma_0$ and $\Gamma_1$
are piecewise linear on a scale $n^{-\frac{1}{2}}$. On the scale of 
order $1$ the incipient piecewise parabolic structure of $\Gamma_1$ 
is discernible.}} 
\label{Fig2}
\end{figure}

We begin by considering
an $n$-sided cell with $n$ very large.
Cells with very many sides are extremely
rare, but {\it if\,} one occurs, then its environment
must look as depicted in figure \ref{Fig2},
where the 
$n$-sided cell of a ``central'' seed at $O$
is surrounded by $n$ 
strongly elongated first neighbor cells containing seeds $F_i$.
Independent evidence for such a geometry comes from work by 
Lauritsen {\it et al.} \cite{LMH93}, 
who Monte Carlo simulated a Hamiltonian favoring the
appearance of many-sided cells. 
In figure \ref{Fig2} four different length scales \cite{HJHletter05} 
play a role, each proportional to its own characteristic power of $n$:

(i) The perimeter $\Gamma_0$ of the central $n$-sided cell 
typically runs within 
an annulus of center $O$, of radius $R_{\rm c} =
(n/4\pi\lambda)^{\frac{1}{2}}$, and of width of 
order $1$. Hence, provided it remains smooth, the perimeter tends
for $n\to\infty$ towards a circle of radius $R_{\rm c}$.
Similarly, the first neighbor seeds $F_i$ will be 
on a circle of radius $2R_{\rm c}$. 

(ii) Sufficient smoothness of the perimeter $\Gamma_0$ 
in the limit $n\to\infty$ is guaranteed by the following property:
locally the radial coordinates of the individual vertices of $\Gamma_0$
have rms deviations of order $n^{-\frac{3}{2}}$ with respect to the
locally averaged radial coordinate.   
A similar statement holds for the curve, not drawn in the figure \ref{Fig2},
that links the successive first neighbors.

(iii) The $n$ vertices 
on the perimeter of the central cell have a line density
$\rho_{\rm vert}=n/(2\pi R_{\rm c})=(n\lambda/\pi)^{\frac{1}{2}}$. 
Consequently, the first neighbor seeds 
$F_i$ have a line density $\frac{1}{2}\rho_{\rm vert}$.

(iv) In the region to the right of the heavy solid line 
$\Gamma_1$, which is occupied by 
second and further neighbors, the 
seed density keeps its ``background'' value $\lambda$,
{\it i.e.} is of order $n^0$ \cite{footnote1}. 
\vspace{1mm}

On this picture we base the following line of arguments.
For $n\to\infty$ the vertex line density 
$\rho_{\rm vert}=(n\lambda/\pi)^\frac{1}{2}$ tends to infinity. 
In spite of this the areal density $\lambda$ of the seeds 
to the right of $\Gamma_1$ stays of order $n^0$. 
It follows that the central cell will have
$\sim n^{\frac{1}{2}}$ second neighbor cells. Since there are $n$ first
neighbors, each second neighbor cell $S_j$ must  
be adjacent to $\sim n^{\frac{1}{2}}$ first neighbor cells $F_i$
(where we call the cells by the names of their seeds and where the symbol
$\sim$ denotes asymptotic proportionality).
Figure \ref{Fig2} shows that under these geometrical constraints
each first neighbor $F_i$ is most likely to have itself four neighbors, 
{\it viz.}~the central $n$-sided cell, a single second neighbor cell, 
and two other first neighbors, $F_{i-1}$ and $F_{i+1}$.

We now focus on the exceptional $F_i$ that have {\it five} neighbors
due to their being adjacent to {\it two\,} second neighbors 
$S_j$ and $S_{j+1}$. 
An example is the cell marked $F_1$ 
in figure \ref{Fig2}, which is adjacent to both $S_1$ and $S_2$.
Let us denote by $f_5$ the fraction of first neighbors that are five-sided.
In view of the scaling relations that precede we expect that 
$f_5=c n^{-{\frac{1}{2}}}+\ldots$\,, where $\,c$ is a numerical
coefficient and the dots indicate terms of higher order in $n^{-\frac{1}{2}}$. 
Any six- and higher-sided $F_i$ will contribute only to these dot terms. 
Hence we have
\bea
m_n &=& 4\,(1-f_5) + 5f_5\nonumber\\[2mm]
&=& 4 + c n^{-{\frac{1}{2}}} + \ldots\,.
\label{relmnc}
\eea
\begin{figure}
\vspace{5mm}
\scalebox{.45}
{\includegraphics{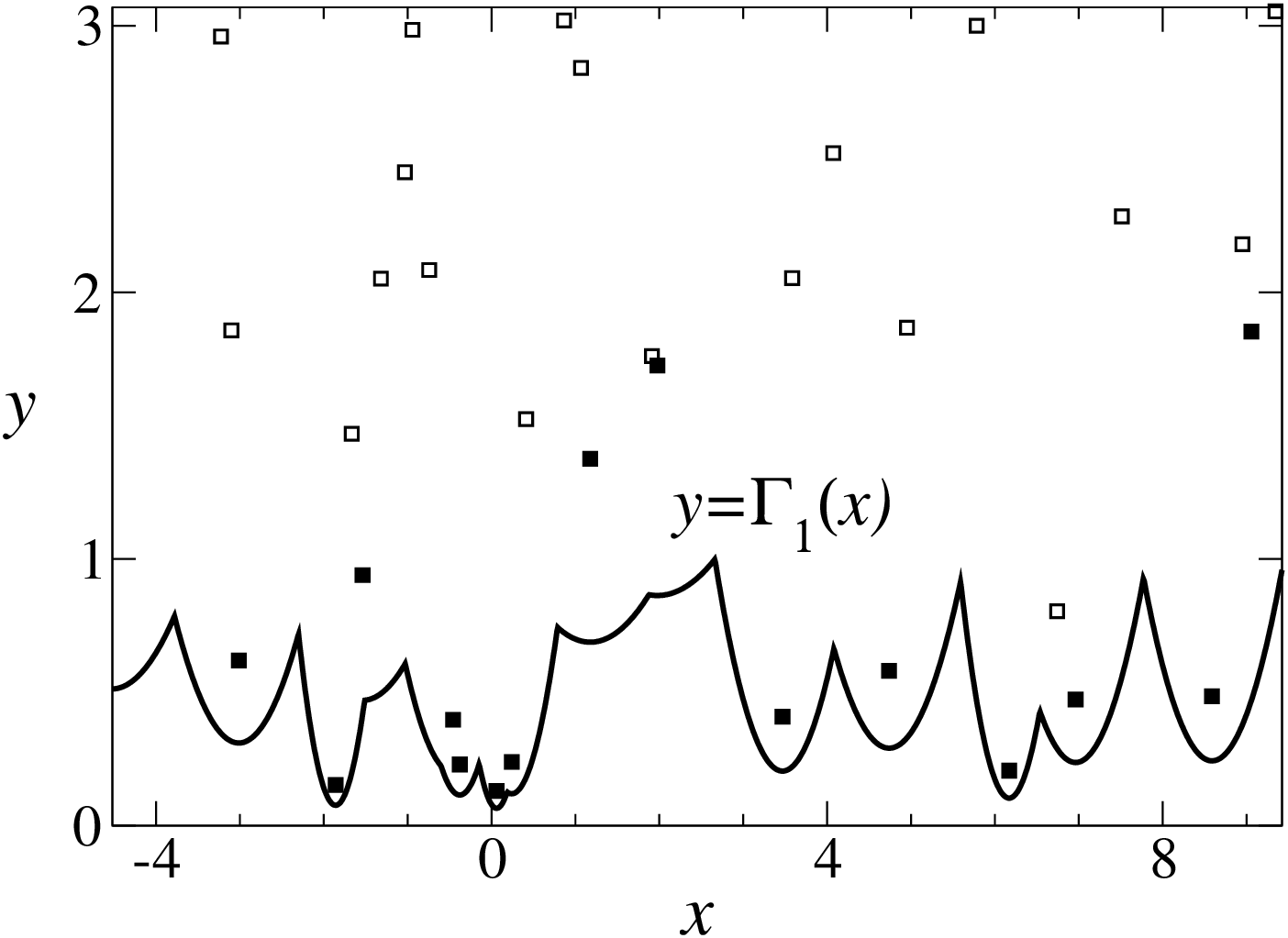}}
\caption{\small The $x$-axis represents a continuum of first neighbors.
    The upper half-plane is randomly filled with seeds of
    uniform density $\lambda$ (here $\lambda=1$).     
    The region of the half-plane which is
    closer to the $x$ axis than to any of the seeds is separated from its
    complement by the piecewise parabolic curve $y=\Gamma_1(x)$.
    The abscissae of the cusps of $\Gamma_1$ 
    are shown in appendix A to have a density of
    $\frac{3}{2}\lambda^{\frac{1}{2}}$ on the $x$ axis. 
    Full and open squares represent second and further neighbors,
    respectively. Note that the horizontal and vertical scales are different.}
\label{Fig3}
\end{figure}
\noindent
It turns out to be possible to determine the constant $c$. To that end
we consider the solid line $\Gamma_1$ in figure \ref{Fig2}, 
which separates the central seed's first from its second neighbors.
In the large-$n$ limit the curve linking the $F_i$
becomes a circle which may locally be replaced with a straight line.
In figure \ref{Fig3} this straight line is represented by the $x$ axis 
and the region of space containing the
second and further neighbors by the half-plane $y>0$.
Seeds are uniformly distributed in the
upper half-plane with the background density $\lambda$.
Since the first neighbors $F_i$ are dense on the $x$ axis,
the curve $\Gamma_1$
divides the half-plane $y>0$ into a lower part 
of points closer to the $x$ axis than to any of the 
seeds, and its complement. 
Hence the function $y=\Gamma_1(x)$ is piecewise parabolic;
its incipient parabolic segments are discernable in figure \ref{Fig2}.
To each cusp of $\Gamma_1(x)$ corresponds a five-sided first neighbor cell. 
Let $\rho_{\rm cusp}$ be the density on the $x$ axis of the abscissae
of the cusps of $\Gamma_1(x)$. 
Then it is clear that $f_5=2\rho_{\rm cusp}/\rho_{\rm vert}$. 
The determination of $\rho_{\rm cusp}$ for given seed density $\lambda$ in
the upper half plane is a well-defined problem in statistics.
A somewhat lengthy calculation 
yields $\rho_{\rm cusp}=\frac{3}{2}\lambda^{\frac{1}{2}}$
(see appendix \ref{secappendix}), in which the coefficient $\frac{3}{2}$ is
essential. 
Using the expression for $\rho_{\rm vert}$ found above we
therefore have that $f_5=2(\pi/n\lambda)^{\frac{1}{2}} \times
3\lambda^{\frac{1}{2}}/2=3(\pi/n)^{\frac{1}{2}}$, 
whence $c=3\pi^{\frac{1}{2}}$.
Substituting this in (\ref{relmnc}) we conclude that for
the Random Voronoi Froth $m_n$ is given by
\beq
m_n=4+3\sqrt{\frac{\pi}{n}}+\ldots, \qquad  (n\to\infty),
\label{mnaspt}
\eeq
where the dots stand for higher order terms in $n^{-\frac{1}{2}}$.
This exact asymptotic formula for the first-neighbor
correlation of Poisson-Voronoi cells
is the key result of this work.
It represents a new
paradigm for the behavior of the cell-cell correlation.
We defer all further comments to section \ref{secdiscussion}.


\subsection{Second-neighbor correlation}
\label{secsecondneighbor}

It is easy to obtain a corollary involving the correlation between 
a central cell and its {\it second\,} neighbor cells in the RVF. 
In a wider context concentric layers of 
cells around a central one
have been the subject of various investigations in recent years
\cite{Asteetal96}.
We denote by $K_n^{(2)}$ the average number of second neighbors
of an $n$-sided cell.
The preceding result allows us to determine almost 
immediately the expression for $K_n^{(2)}$, again in the limit of 
asymptotically large $n$.
We appeal once more to figure \ref{Fig2}.
For $n\to\infty$ the circle of seeds $F_i$ 
has a circumference $4\pi R_{\rm c}=2(n\pi/\lambda)^{\frac{1}{2}}$.
The total number $N_{\rm cusp}$ of cusps of $\Gamma_1$ 
along this circle is equal to $\rho_{\rm cusp}$ times the circumference, 
whence $N_{\rm cusp}=3(\pi n)^{\frac{1}{2}}$.
Some caution must be exercised at this point. 
The number of second-neighbor cells can
at most be equal to $N_{\rm cusp}$; however, it will be slightly smaller.
The reason is that the common border between
a given second neighbor and the set of first neighbors 
may consist of more than a single parabolic segment.
We therefore have $K_n^{(2)}=\eta N_{\rm cusp}$ where
the reduction factor $\eta<1$ is a geometrically defined 
mathematical constant. The result is that
\beq
K_n^{(2)} = 3\eta(\pi n)^{\frac{1}{2}} + \ldots, \qquad (n\to\infty).
\label{K2naspt}
\eeq
We have not seen a way to find a simple analytic expression for $\eta$;
from a simulation in which we generated more than
2000 cusps we obtained the estimate  
\beq
\eta= 0.95 \pm 0.01.
\label{numeta}
\eeq
Equation (\ref{K2naspt}) together with (\ref{numeta})
constitutes the exact asymptotic formula for the total number of
second neighbors of a central $n$-sided cell 
in a Random Voronoi Froth.


\section{Discussion}
\label{secdiscussion}


We now discuss the significance of equation (\ref{mnaspt}).
This equation is first of all a statement of principle:
in the RVF 
(and hence in all theoretical models or experimental systems
systems for which it is relevant),
as the accuracy and the range of the data increase,
the two-cell correlation 
$m_n$ should be seen to follow an $n^{-\frac{1}{2}}$ law
and asymptotically approach $m_\infty=4$ \cite{footnote2}.
Equation (\ref{mnaspt}) therefore rules out
the possibility of the existence of a general
proof of Aboav's law for an unrestricted range of $n$.
Aboav's law keeps its meaning, however, as a good
linear approximation to the available data
in the window accessible to simulations or experiments.

\subsection{Comparison to RVF simulation data}
\label{subcomparison1}

\begin{figure}
\vspace{5mm}
\scalebox{.50}
{\includegraphics{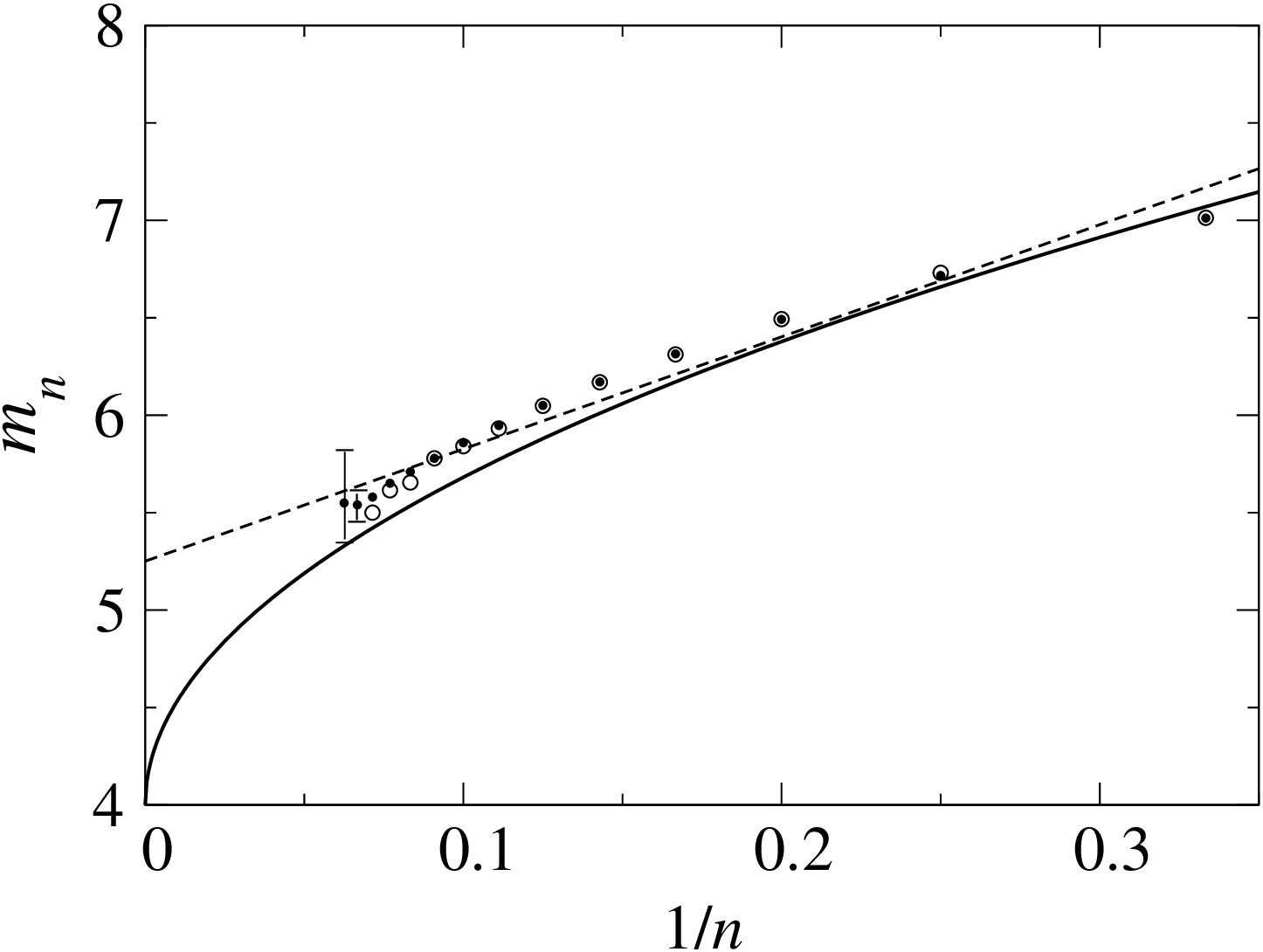}}
\caption{\small 
     Average sidedness $m_n$ of a neighbor of an $n$-sided cell
     plotted against $n^{-1}$.
     Open circles and closed dots: same data as in figure
     \ref{Fig1}, but in this plot the curvature appears more clearly.
     Dashed straight line: Aboav's 
     law $m_n=a+bn^{-1}$ with $a$ and $b$ as in figure \ref{Fig1}.
     Solid line: first two terms, $m_n=4+3(\pi/n)^{\frac{1}{2}}$,
     of the exact asymptotic series (\ref{mnaspt}). Error bars are shown for
     the dotted data; where absent, they are smaller than the data points. }  
\label{Fig4}
\end{figure}

Earnshaw and Robinson \cite{EarnshawRobinson94}, following a remark
for which they credit Weaire, have emphasized that in order to detect
small deviations from Aboav's law it is essential
to plot the $m_n$ data as a function of $n^{-1}$.
We fully concur with them and have plotted in this way in figure 
\ref{Fig4} the same data that were shown in figure \ref{Fig1}.
In the new variable $n^{-1}$ Aboav's law remains linear and
the negative curvature of the simulation data now appears clearly.
In addition we have represented by a solid line in figure \ref{Fig4} 
the first two terms,
$m_n=4+3(\pi/n)^{\frac{1}{2}}$, of equation (\ref{mnaspt}).
We discuss separately two aspects of the theory, {\it viz.} its prediction of
the curvature effect and its numerical accuracy.

{\it Curvature.\,\,--}
We confront our calculation of the asymptotic behavior of $m_n$ with
the RVF simulation data. 
It is visually apparent from figure \ref{Fig4} that the 
simulation data have a curvature very similar to that of the 
theoretical curve. Quantitatively, we obtain from 
(\ref{mnaspt}) analytically
\beq
-\frac{(nm_n)''}{nm_n}=
\frac{ 3\pi^{\frac{1}{2}} }{ 4n^{\frac{5}{2}} [4+3(\pi/n)^{\frac{1}{2}}] }\,,
\eeq
which ranges from $0.006$ to $0.001$ in the interval $4\leq n\leq 8$
and hence is of the same order of magnitude as the value $0.004$
found from figure \ref{Fig1}.
We conclude that the asymptotic result (\ref{mnaspt})
not only explains why the $m_n$ data should be curved, but correctly
predicts the sign and magnitude of the curvature.

{\it Numerical accuracy.\,\,--}
Equation (\ref{mnaspt}) is the beginning of an asymptotic series 
in powers of $n^{-\frac{1}{2}}$.
Although they may, such series are not guaranteed
to produce accurate numerical values when the expansion
variable is finite. 
Figures \ref{Fig2} (which is for $n\approx 100$) and \ref{Fig3} 
(for $n=\infty$), used in our derivation,
refer to seed configurations that 
are very far away indeed from those observed in
either simulations or experiments. Therefore,
one might be uninclined {\it a priori} 
to expect of this expansion a high degree of numerical accuracy.
However, figure \ref{Fig4} shows the contrary.
When compared to the numerical RVF values,
our asymptotic result (\ref{mnaspt}), limited to its first two terms,
is quite good: it is only slightly more off than Aboav's linear fit.
This gives confidence in the expansion not only as an asymptotic 
constraint on the $m_n$ curve, but also as a practical tool
for estimating this and other correlations.
We will present further numerical considerations
in section \ref{secnumerical}.


\subsection{Comparison to diffusion limited colloidal aggregation (DLCA)}
\label{seccomparison2}

The RVF is an idealized model which appears in the discussion
and analysis of various naturally occurring 
cellular systems \cite{Rivier93}.
We now investigate the relevance 
of the RVF cell-cell correlations calculated in section
\ref{sectwocellcorrelations}
for a real physical system, {\it viz.}~diffusion limited colloidal
aggregation (DLCA) in two-dimensional suspensions.
In the experiments one monitors the slow aggregation 
of particles trapped at the air/water interface and initially randomly
distributed. 
In the early stage of this process isolated clusters appear and
video images of the system taken after some time $t$ are
analyzed in terms of the Voronoi cells constructed around the centers 
of mass of these clusters.
Simulation methods developed for such aggregating
systems are known \cite{EHR96} to
agree well the experimental studies 
and we will discuss the simulations first.

\begin{figure}
\vspace{5mm}
\scalebox{.50}
{\includegraphics{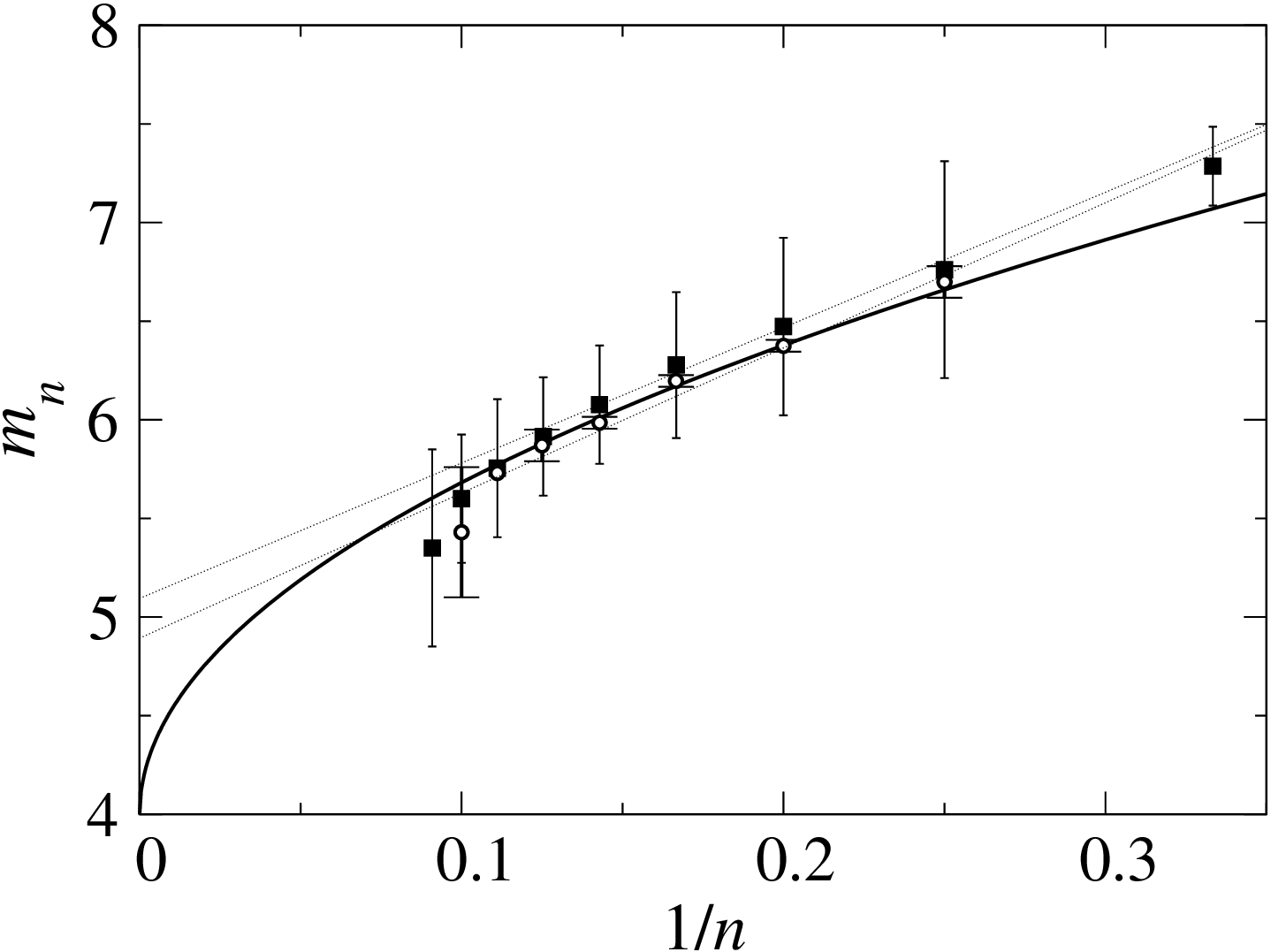}}
\caption{\small 
   Average sidedness $m_n$ of a neighbor of an $n$-sided cell.
   Full squares: DLCA simulation data by Fern\'andez-Toledano {\it et al.}
   (reference \cite{FTetal05}, inset of figure 5; the error bars were kindly
   provided to us by the authors). Open circles: experimental DLCA data by
   Earnshaw and Robinson (data for $t=60$ min
   from \cite{EarnshawRobinson94}, figure 5).
   Thin straight lines: Aboav's linear law; the upper one is the fit to the
   simulation data and the lower one the fit to
   the experimental data proposed in reference \cite{FTetal05}
   and in reference \cite{EarnshawRobinson94,EarnshawRobinson95},
   respectively.
   Solid curve: 
   first two terms of our theoretical large-$n$ expansion
   for the Random Voronoi Froth, equation (\ref{mnaspt}).} 
\label{Fig5}
\end{figure}

Recently detailed Brownian dynamics simulations
were performed by Fern\'andez-To\-le\-da\-no {\it et al.}~\cite{FTetal05}.
Their $m_n$ {\it versus\,} $1/n$ data
have been represented in our figure \ref{Fig5}.
The authors fit the $m_n$ data by Aboav's linear law
(upper thin straight line in figure \ref{Fig5}), 
but their data points depart from linearity for $n>9$ 
in an even more pronounced way than those of figure \ref{Fig4}.
Fern\'andez-Toledano {\it et al.}
attribute these deviations to statistical uncertainties. 
Instead, we believe that their simulations were
accurate enough for them to actually see the curvature
predicted by our asymptotic RVF theory,
represented in figure \ref{Fig5} by the solid curve. 
It predicts a curvature of the same order of
magnitude as observed in the data.
While both Aboav's law and our curve 
are compatible with the error bars,
our theory provides an explanation whereas the linear fit does not.
\begin{figure}
\vspace{5mm}
\scalebox{.50}
{\includegraphics{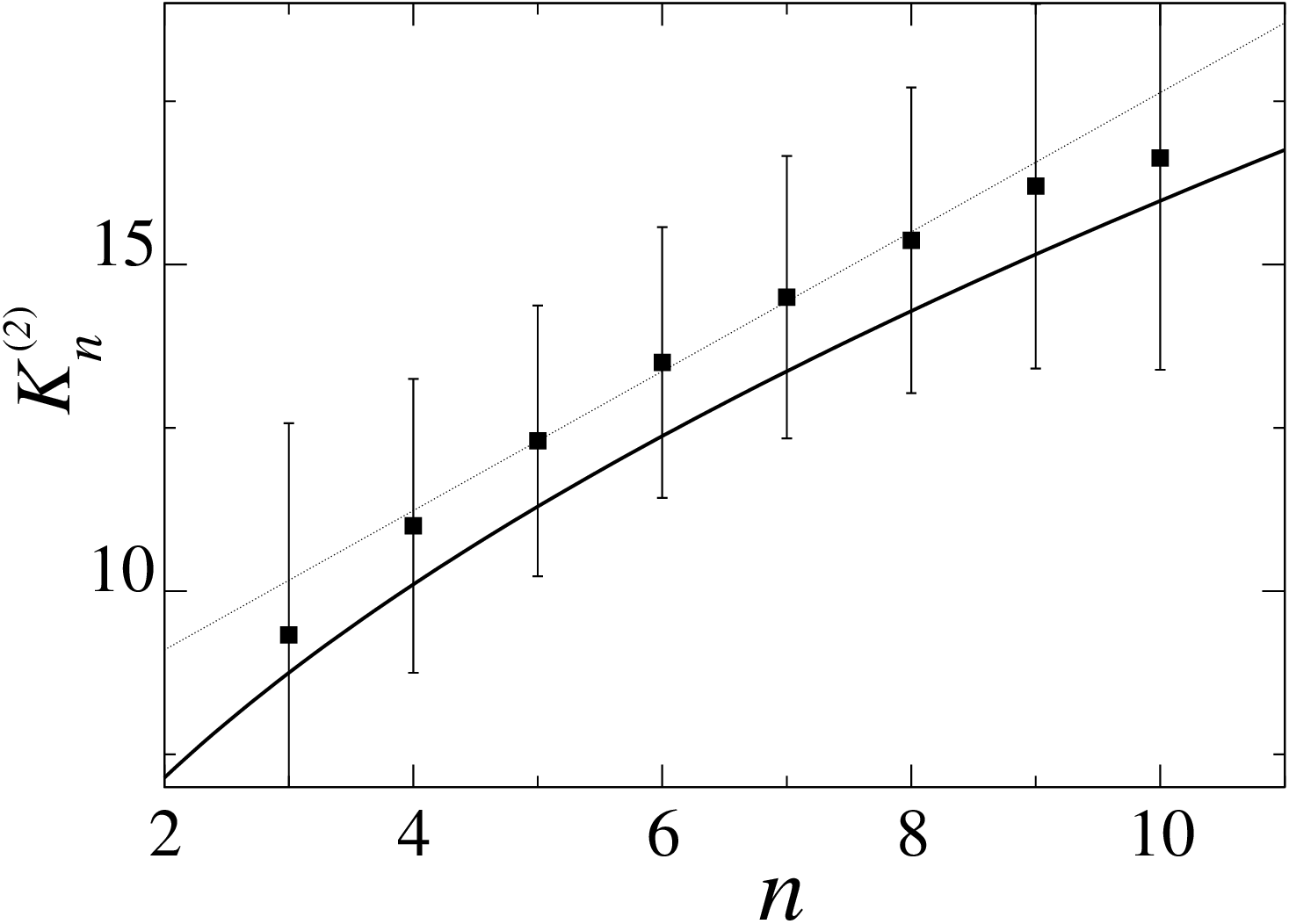}}
\caption{\small 
    Average number $K^{(2)}_n$ of second neighbors of an $n$-sided cell.
    Full squares: DLCA simulation data by Fern\'andez-Toledano {\it et al.}
    (from \cite{FTetal05}, figure 7; the error bars were kindly provided
    to us by the authors).  Thin straight line: the linear fit to the data
    proposed in reference \cite{FTetal05}.
    Solid curve: first term of our theoretical large-$n$
    expansion for the Random Voronoi Froth, equation (\ref{K2naspt}) with
    $\eta=0.95$.} 
\label{Fig6}
\end{figure}
The authors of reference \cite{FTetal05} 
fit their $K_n^{(2)}$ data, shown in figure \ref{Fig6}, 
with a linear curve
based on Aboav's law plus some additional hypotheses.
The data show, however, a 
slight but clearly distinguishable curvature
of the same sign and magnitude as predicted by our law (\ref{K2naspt});
hence we believe they have seen this law.

The simulations were prompted by experimental DLCA studies 
carried out in the nineties by Earnshaw and Robinson,
who among several other questions also tested 
\cite{EarnshawRobinson94,EarnshawRobinson95}
the validity of Aboav's law.
Earnshaw and Robinson fitted their experimental 
$m_n$ {\it versus} $1/n$ data by a straight line
and concluded \cite{EarnshawRobinson94} that 
DLCA accords well with the Aboav-Weaire law.
We show one of their data sets in our figure \ref{Fig5}. 
It exhibits the downward curvature that we expect.
Our comment is again that whereas within the error bars
a linear fit is also compatible with the data,
only the curved theoretical law (\ref{mnaspt}) provides an explanation.
The experimental data seem only a small step away
from being able to decisively rule out either one or the other.
In view of the agreement observed by the authors between
experimental and simulation data in many other respects \cite{EHR96},
we are confident that DLCA experiments pushed to greater resolution
will make the curvature visible. 

We finally ask what the physical reason is
why RVF theory is applicable to DLCA. We believe the answer is as follows.
One may obtain the Voronoi construction for randomly
distributed seeds by starting a circular domain growth
simultaneously from each seed and having it stop at the contact
points of the circular frontiers.
It is likely that in the experiments the depletion zones arising around
each cluster play exactly the role of these growing circles.


\subsection{Comparison to other exact models}
\label{seccomparison3}

There are very few exact results 
for the two-cell correlation $m_n$ in other geometrical systems.
Those that we are aware of all concern model systems artificially 
constructed to be exactly solvable. 
An example is the anisotropic ``laminated Poisson
network'' studied by Fortes \cite{Fortes95}, for which Aboav's law appears to
be satisfied exactly (for $n\geq 4$) with $a=\frac{4}{3}$ and $b=14$.
Several results do exist, however, for {\it to\-po\-lo\-gi\-cal} 
planar cell models, {\it i.e.} models in which only the vertex 
connectivity is considered but where metric properties such as distances 
and areas are ignored.
To this class belongs very interesting work by Le Ca\"er \cite{LeCaer91}
and Le Ca\"er and Delannay \cite{LeCaerDelannay93}. These authors
start from regular planar $z$-coordinated lattices with $z>3$ and 
by a suitable algorithm construct from it a random topological froth
that is three-coordinated just like the RVF
and the vast majority of other cellular systems.
For one of the simplest models of this kind (the case $z=4$)
Le Ca\"er \cite{LeCaer91} finds
\beq
m_n = \frac{9}{2} + \frac{10}{n}, \qquad (n=4,5,...,8),
\label{eqnmnLeCaer}
\eeq
which is Aboav's law on a restricted $n$ interval.
Other topological models studied by these authors
may be treated numerically exactly
and, in the general case, show deviations from Aboav's law.
 
Also in the class of topological models is work by
Godr\`eche {\it et al.} \cite{GKY92}, who studied
an ensemble of diagrams encountered in field theory.
For these they found
\beq
m_n=7+\frac{3}{n}+\frac{9}{n(n+1)},
\label{eqnmnGKY}
\eeq
which while curving upward approaches
asymptotically the linear law $nm_n\simeq 7n$.

It may well prove useful to classify models of cellular structure 
according to the decay of their $m_n$ by writing
generically $m_n-m_\infty \sim n^{-\alpha}$, where $\alpha$ is a positive
exponent. 
Equations (\ref{mnaspt}) and (\ref{eqnmnGKY}) then provide
examples with $\alpha=\frac{1}{2}$ and $\alpha=1$, respectively.
Beyond the RVF studied above
it is unknown at present which other theoretical models and
experimental systems have $\alpha=\frac{1}{2}$, but for DLCA
this certainly seems the likely value.
We do not know if it is conceivable that a microscopic {\it geometrical\,}
model of cellular structure, as opposed to the {\it topological\,} ones,
could have $\alpha=1$. 
If so, its $nm_n$ would obey Aboav's linear law, 
either exactly for all $n$ or asymptotically for large $n$. 
Obviously, knowing a system's $\alpha$ value 
is relevant for the analysis of
its data. This value is very likely tied up with the cell 
formation mechanism, but in an as yet completely unknown way.


\section{Beyond exact asymptotics: curve fitting}
\label{secnumerical}

Equation (\ref{mnaspt}) is an exact expansion in powers of 
$n^{-\frac{1}{2}}$ that we may write as
\beq
m_n = 4 + \frac{ 3\pi^{\frac{1}{2}} }{ n^{\frac{1}{2}} }   
+ \frac{a_2}{ n^{\!\!\!\phantom{\frac{1}{1}}} }
+ \frac{a_3}{ n^{\frac{3}{2}} } 
+\ldots
\label{mnfullaspt}
\eeq
The higher order coefficients $a_2, a_3, \ldots$ 
are well-defined and it is possible, in principle, 
to determine them successively.
It would certainly be interesting to know a few more of them,
but the corresponding calculations are far from 
straightforward. Moreover, it should be kept in mind that
the asymptotic series (\ref{mnfullaspt}) is in all likelihood
{\it only} asymptotic, 
which means (see {\it e.g.} \cite{BH86}, chapter 1)
that even though the first few terms may give a good approximation, 
when adding more and more terms for fixed $n$ one eventually finds that
the series does not converge. 
All we know is that as $n$ grows,
(\ref{mnfullaspt}) approaches the exact result ever more closely. 

The remainder of this section is in a spirit different 
from the rest of our work. We include it because of the interest in
numerical fits that exists in part of the cellular system community 
(see, {\it e.g.,} reference \cite{Okabeetal00}).
In the same pragmatic way that 
led to Aboav's law we now ask if it is possible to
fit the numerical $m_n$ data
by a simple analytic expression that
{\it incorporates the first two terms of the 
asymptotic constraint} (\ref{mnaspt}).
The natural answer is to 
extend the two explicitly known terms of (\ref{mnaspt})
by one or more extra terms with coefficients chosen to fit the data,
and to truncate after that.
With two extra terms one gets
\beq
m_n = 4 + 3\pi^{\frac{1}{2}}  n^{-\frac{1}{2}}  + 
A n^{-1} + B n^{-\frac{3}{2}},  
\label{hjhfit2}
\eeq
where $A$ and $B$ are on the same footing as
Aboav's parameters $a$ and $b$.
With three extra terms one has
\beq
m_n = 4 + 3\pi^{\frac{1}{2}}  n^{-\frac{1}{2}}  + 
A n^{-1} + B n^{-\frac{3}{2}} +Cn^{-2}. 
\label{hjhfit3}
\eeq
Here $A,B,C,\ldots$ are not related in any simple way to 
$a_1,a_2,a_3,\ldots$ of equation (\ref{mnfullaspt}).
A relation between the fit parameters arises
if one imposes that the fit obey 
Weaire's sum rule \cite{Weaire74}
\beq
\langle n m_n \rangle =\langle n^2 \rangle,
\label{weaire}
\eeq
in which $\langle X_n \rangle \equiv \sum_{n=3}^\infty X_n p_n$
for any function $X_n$ of the sidedness $n$,
and $p_n$ is the probability that an arbitrarily chosen cell 
have $n$ sides. 
Thus, combining (\ref{weaire}) with Aboav's law (\ref{eqnmn})
leads to the well-known relation $b=\mu_2+6a$ where
$\mu_2\equiv\langle n^2 \rangle - \langle n \rangle^2$ and one uses that 
$\langle n \rangle=6$.
Imposing Weaire's sum rule 
makes good sense and we will therefore do the same.
After substituting the assumed expressions (\ref{hjhfit2}) and (\ref{hjhfit3})
in (\ref{weaire}) we find 
\beq
A + B \langle n^{ -\frac{1}{2} } \rangle 
= 12 + \mu_2 - 3\pi^\frac{1}{2} \langle n^\frac{1}{2} \rangle.
\label{rel2}
\eeq
and
\beq
A + B \langle n^{ -\frac{1}{2} } \rangle + C \langle n^{-1} \rangle
= 12 + \mu_2 - 3\pi^\frac{1}{2} \langle n^\frac{1}{2} \rangle,
\label{rel3}
\eeq
respectively.
The numerically most accurate data
for $p_n$ and $m_n$ available today are those due to Brakke \cite{Brakke}.
They lead to $\mu_2=1.7807$,
$\langle n^\frac{1}{2} \rangle=2.4344$,
$\langle n^{-\frac{1}{2}} \rangle=0.4160$,
and $\langle n^{-1} \rangle=0.1753$,
which are all needed in (\ref{rel2}) and (\ref{rel3}).

We fitted both expressions (\ref{hjhfit2}) and (\ref{hjhfit3})
by minimizing the weighted maximum deviation from the data, {\it i.e.,} 
letting $m_n^{\rm B}$ denote Brakke's numerical values and $\sigma_n^{\rm B}$
their standard deviation, by searching for 
\beq
\epsilon = \min\,\,\max_{3 \leq n \leq 15} 
|m_n - m_n^{\rm B}|/\sigma_n^{\rm B},
\label{minmaxfit}
\eeq
where the minimum is taken over the fit parameters 
under either constraint (\ref{rel2}) or (\ref{rel3}). 
The results are shown in Table I. 
As witnessed by the value of $\epsilon$, 
the two-parameter fit still runs far
outside of the standard deviations given by Brakke.
The three-parameter fit 
yields strongly improved results.
Comparison of the two- and three-parameter fit
shows, in particular, that as more terms are added, there is no sign
that the values of $A,B,\ldots$ converge.
This also is a strong indication of the merely 
asymptotic character of the $n^{-\frac{1}{2}}$ expansion.

We emphasize that whereas sections \ref{sectwocellcorrelations} and 
\ref{secdiscussion} of our work
rest on first principles, the procedure 
followed in the present section does not. 
It is hybrid in that it represents the construction of
a best fit which has been made to include 
the first two terms of the exact large-$n$ expansion. 
There is of course no reason to believe that $m_n$ can be represented 
as a sum of a finite number of powers of $n$, whether it be 
(\ref{hjhfit2}), (\ref{hjhfit3}), or Aboav's law (\ref{eqnmn}).
\begin{table}
\begin{center}
\renewcommand{\arraystretch}{1.25}
\begin{tabular}[t]{||r |l r@{.}l |l r@{.}l |l@{$\,\pm\,$}l ||}
\hline
&\multicolumn{3}{|c|} {Two-parameter fit}
&\multicolumn{3}{|c|} {Three-parameter fit} 
&\multicolumn{2}{|c||} {Simulation data}   \\
&\multicolumn{3}{|c|} {$A=\phantom{-}3.816$ }
&\multicolumn{3}{|c|} {$A=\phantom{-1}6.279$ } 
&\multicolumn{2}{|c||} {(Brakke \cite{Brakke})}  \\
&\multicolumn{3}{|c|} {$B=-7.163$ }
&\multicolumn{3}{|c|} {$B=-18.327$ } 
&\multicolumn{2}{|c||} {} \\
&\multicolumn{3}{|c|} {}
&\multicolumn{3}{|c|} {$C=\phantom{-}12.440$ } 
&\multicolumn{2}{|c||} {}  \\
&\multicolumn{3}{|c|} {$\epsilon=50.46$}
&\multicolumn{3}{|c|} {$\epsilon=\phantom{-1}7.013$ } 
&\multicolumn{2}{|c||} {}  \\
\hline       
$n$  
&\multicolumn{1}{|c}{ $m_n$ } 
&\multicolumn{2}{c|}{ $m_n-m_n^{\rm B}$ } 
&\multicolumn{1}{|c}{ $m_n$ } 
&\multicolumn{2}{c|}{ $m_n-m_n^{\rm B}$ } 
&\multicolumn{2}{|c||}{ $m_n^{\rm B}\,\, \pm \,\,\sigma_n^{\rm B}$ } \\
\hline
$3$  & 6.9635   & -0&0481  & 7.0183  &  0&0067  & 7.0116   & $ 0.0095$ \\
$4$  & 6.7173   &  0&0002  & 6.7152  & -0&0019  & 6.7171   & $ 0.0028$ \\
$5$  & 6.5005   &  0&0083  & 6.4923  &  0&0001  & 6.4922   & $ 0.0017$ \\
$6$  & 6.31943  &  0&00442 & 6.31593 &  0&00092 & 6.31501  & $ 0.00014$ \\
$7$  & 6.16816  & -0&00256 & 6.17114 &  0&00042 & 6.17072  & $ 0.00016$ \\
$8$  & 6.04042  & -0&00974 & 6.04932 & -0&00084 & 6.05016  & $ 0.00021$ \\
$9$  & 5.93116  & -0&01622 & 5.94497 & -0&00242 & 5.94739  & $ 0.00035$ \\
$10$ & 5.83659  & -0&02200 & 5.85429 & -0&00431 & 5.85860  & $ 0.00064$ \\
$11$ & 5.7538   & -0&0254  & 5.7746  & -0&0047  & 5.7793   & $ 0.0014$ \\
$12$ & 5.6807   & -0&0297  & 5.7038  & -0&0066  & 5.7104   & $ 0.0082$ \\
$13$ & 5.6155   & -0&0364  & 5.6404  & -0&0115  & 5.6519   & $ 0.0082$ \\
$14$ & 5.557    & -0&0261  & 5.583   &  0&000   & 5.583    & $ 0.023$ \\
$15$ & 5.504    & -0&0310  & 5.531   & -0&004   & 5.535    & $ 0.070$ \\
\hline
\end{tabular}
\caption{\normalsize  Two-parameter fit (\ref{hjhfit2}) and three-parameter
fit (\ref{hjhfit3}) to Brakke's simulation data \cite{Brakke}. 
The latter are denoted here
as $m_n^{\rm B}$ and their statistical error as $\sigma_n^{\rm B}$.
Both fits have been made to satisfy Weaire's sum rule (\ref{weaire}).}
\label{TableI}
\end{center}
\end{table}


\section{Theory for non-RVF systems}
\label{secnonRVF}

The RVF serves as 
a model of reference for general cellular systems
much in the same way as an ideal gas does for interacting gases:
some of the RVF result will apply to other systems and some will not.
There are many factors that may potentially
cause an experimental system
to depart from RVF-like behavior. 
{\it E.g.,}
(i) in some cellular structures 
the `seeds' represent actual particles or larger
physical entities whose mutual interactions cannot be neglected;
(ii) a general planar cell structure cannot be derived from a set
of point centers by means of the Voronoi construction; and
(iii) some systems, like soap froths, are not in equilibrium but rather
in a coarsening state and need a dynamical theory in terms of cell
transformation processes.
We refer to Rivier \cite{Rivier93} for an overview.

We now ask the subtler question as to whether the cell-cell correlation 
$m_n$ in any of these non-RVF systems also deviates from Aboav's law by
some slight curvature, and if it does, what the
asymptotic behavior of its $m_n$ is.
For certain microscopic models one may hope
to be able to find the answer to this question perturbatively
starting from the analysis of the present work.

To show how such an approach might proceed
we become more speculative.
We discuss briefly and heuristically
an example from class (ii) above, 
{\it viz.} the Voronoi tessellation
associated with a gas of hard core particles of finite diameter $d$.
For this system the preceding analysis of the large $n$ limit
remains valid as long as $n \lesssim n^*$, where
the crossover value $n^*$ is 
determined by the condition that the distance between adjacent 
first neighbors become comparable to the particle diameter. This gives
$2\pi R_{\rm c}/n^* \sim d$ whence, because of 
$R_{\rm c}=(n/4\pi\lambda)^{\frac{1}{2}}$,
it follows that $n^* \sim 1/(\lambda d^2)$,
with the $\sim$ sign indicating asymptotic proportionality.
For $n \gsim n^*$ the repulsion between the particles combined with the
condition that they be locally aligned along a circle
imposes that the radius of the central cell grows as $R_{\rm c} \sim nd$
and hence that $f_5$ must saturate at a value
$f_5 = c_0 \lambda^{\frac{1}{2}} d$ where $c_0$ is an unknown numerical
coefficient. 
The correlation $m_n$ can depend only on the two dimensionless 
variables $n$ and $\lambda d^2$.
It is reasonable to assume that in the scaling limit
$n\to\infty, \lambda d^2\to 0$ with $ n\lambda d^2=x$ fixed,
$m_n$ may be expressed as
\beq
m_n \simeq 4 + \lambda^{\frac{1}{2}}d {\cal M}(n\lambda d^2) 
\label{scalingmn}
\eeq
in which the scaling function ${\cal M}$ must satisfy
\beq
{\cal M}(\infty)=c_0, \qquad {\cal M}(x) \simeq 
\frac{3\pi^{\frac{1}{2}}}{x^{\frac{1}{2}}}
\mbox{\,\,\, for \,\,\,} x\to 0.
\label{propscalM}
\eeq
Hence for hard core particles the limiting value $m_\infty$
of (\ref{mnaspt}) is changed and (\ref{propscalM}) does not tell us
how this new limit is approached for large $n$. 
However, the $n^{-\frac{1}{2}}$ decay law and the curvature effect that it
entails survive
in the crossover regime $n \lesssim \lambda d^2$,
and this is the regime that is encountered first when the
experimental or simulational precision increases.  
Future work will have to deal with
this and other instances of deviations from RVF statistics.
\vspace{1mm}


\section{Conclusion}
\label{secconclusion}

Aboav's linear law (\ref{eqnmn}) for
the two-cell correlations $m_n$ has been known to fail
for the Random Voronoi Froth.
The exact calculation of this work has explained
both qualitatively and quantitatively
why it must fail: the exact asymptotic formula for $m_n$ 
exhibits an inverse square root decay with $n$.
This large-$n$ behavior represents a new paradigm in the field of
planar cellular systems.

In our discussion
we have raised the question of whether similar violations
occur also in other planar cellular systems, be they  
theoretical models or experimental realizations. 
We found that departures from Aboav's law
are convincingly present in computer simulations of
diffusion limited colloidal aggregation (DLCA) 
performed recently by Fern\'andez-Toledano {\it et al.}~\cite{FTetal05}.
Since DLCA simulations have been found to be in all respects 
close to the corresponding experiments \cite{EHR96},
it is very likely that DLCA experiments 
violate Aboav's linear law in the same way.
Experiments by Earnshaw and Robinson 
\cite{EarnshawRobinson94,EarnshawRobinson95},
in spite of error bars that do not unambiguously
rule out Aboav's law,
provide evidence for the curvature effect.
One must expect that future experimental data 
(or perhaps even reanalysis of existing data), 
provided they are of high enough precision 
and/or cover a large enough range in $n$, 
will reveal the presence  
of curvature in the $m_n$ {\it versus} $1/n$ relation
also for certain 
cellular structures other than DLCA.

The parameters $a$ and $b$ 
of Aboav's law are useful in that they provide
a rough global classification of the behavior of a
planar cellular system in the experimentally accessible
range of measurement. 
No first-principle analysis of any geometrical cell model
has, however, produced an interpretation and theoretical expressions
for these parameters.
Empirically, moreover, the precise values of $a$ and $b$ depend on the
range of the fit and on how it is performed \cite{LeCaer91}.
Therefore in future high resolution simulations or experiments
a fixation on Aboav's law would be misguided and any curvature effect,
whenever there is evidence for it, will be worthy of study.
\vspace{4mm}

{\it Acknowledgment.\,\,--}
This work was made possible by a six month sabbatical period (CRCT)
granted to the author by the French Ministry of Education in the
academic year 2004-2005. The author thanks Dr.~J.C. Fern\'andez-Toledano
for making his error bar data available to him. He also thanks Dr. K.A.
Brakke for correspondence and Dr. N. Rivier for a discussion.
\appendix

\section{Appendix}
\label{secappendix}

Equation (\ref{mnaspt}) is the central result of this paper.
Determining the coefficient of the second term in this equation
involves a problem in statistics that we solve in this appendix.
We have to prove the following theorem, illustrated by figure \ref{Fig3}. 
\vspace{2mm}

\noindent {\sc Theorem.}
{\it Let point particles {\rm(}{\it `seeds'\,}{\rm)} be randomly distributed 
with uniform density $\lambda$ in the upper half plane $y>0$.
Let $\Gamma_1(x)$ be the piecewise parabolic curve that
separates the upper half plane into a region 
of points closer to the $x$ axis
than to any of the seeds, and its complement.
Consider the set of
abscissae of the points where the parabolic segments join and where
therefore $\Gamma_1(x)$ has cusps.
Then the density $\rho$ of the abscissae on the $x$ axis is equal to
$\rho=\frac{3}{2}\lambda^{\frac{1}{2}}$. }
\vspace{2mm}

\noindent {\sc Proof.}
The nontrivial part of this theorem is the coefficient $\frac{3}{2}$,
since the seed density $\lambda$ may be scaled away; below we 
keep $\lambda$  only to have at all times a check on the dimensionality 
of the quantities involved in the calculation.

Let $P_i=(x_i,y_i)$ be the position of the $i$th seed. The parabola
\beq
f_i(x)=\frac{y_i}{2}\Big(1+\frac{(x-x_i)^2}{y_i^2}\Big)
\label{deffix}
\eeq
separates the upper half plane into a region containing all points closer to
the $x$ axis than to $P_i$, and its complement.
Each parabolic segment of $\Gamma_1(x)$ lies on one of the $f_i(x)$.

For the considerations that follow we must, 
as is always implied in statistical mechanics, 
take a finite system and let eventually its size tend to infinity.
We will start with a rectangular box $[-L,L]\times[0,L]$
whose volume we denote by $V=2L^2$ \cite{footnote4}.
Let the seeds inside this box be those of indices 
$i=1,2,\ldots,N$, where $N$ is such that $N/V=\lambda$.

We choose a small $\Delta x$ and
define $\rho\Delta x$ as the probability that there is a cusp with abscissa
in the interval $[0,\Delta x]$. This is the probability
that there exist two parabolas,
say $y=f_j(x)$ and $y=f_k(x)$, that have a
point of intersection $P_{jk}=(x_{jk},y_{jk})$ with 
$x_{jk}\in[0,\Delta x]$ and such that $P_{jk}$ is below
all the parabolas $y=f_\ell(x)$ with $\ell\neq j,k$.
This may be expressed as
\beq
\rho\Delta x=
\int_{-L}^L \prod_{1\leq i\leq N} \frac{\dd x_i}{2L}
\int_0^L \prod_{1\leq i\leq N} \frac{\dd y_i}{L}
\sum_{1\leq j<k\leq N}\chi_{jk}(\Delta x)
\prod_{ \substack{1\leq\ell\leq N \\[1mm] \ell\neq j,k} } 
\theta_{\ell}(y_{jk})
\label{defrDx}
\eeq
where 
\beq
\chi_{jk}(\Delta x)=
\left\{
\begin{array}{ll}
1 & \mbox{ if\, } 0 \leq x_{jk} \leq  \Delta x\\[2mm]
0 & \mbox{ otherwise}
\end{array}
\right.
\eeq
and
\beq
\theta_{\ell}(y_{jk})=
\left\{
\begin{array}{ll}
1 & \mbox{ if\, } f_\ell(x_{jk}) > y_{jk}\\[2mm]
0 & \mbox{ otherwise. }
\end{array}
\right.
\eeq
The sum in the integrand in (\ref{defrDx}) 
is on $\frac{1}{2}N(N-1)$ terms that all
give identical results so that we may choose $j,k=1,2$.
The $N-2$ integrals on $(x_3,y_3),\ldots,(x_N,y_N)$ 
may be carried out independently; we have $\theta_\ell(y_{12})=1$ 
unless $(x_\ell,y_\ell)$ is in the disk of radius $y_{12}$ centered at
$(x_{12},y_{12})$. Hence 
\beq
\int_{-L}^L\!\dd x_\ell\int_0^L\!\dd y_\ell\,\theta_\ell(y_{12})
= V-\pi y_{12}^2
\label{intxlyl}
\eeq
Using this we may rewrite (\ref{defrDx}) as
\beq
\rho\Delta x= \frac{N(N-1)}{2V^N}
\int_{-L}^L\!\dd x_1 \int_{-L}^L\!\dd x_2
\int_0^L\!\dd y_1 \int_0^L\!\dd y_2\,\,
\chi_{12}(\Delta x)\,\big[V-\pi y_{12}^2\big]^{N-2}.
\label{eqnrDx1}
\eeq
In the limit $N,V,L\to\infty$ with $N/V=N/(2L^2)=\lambda$ fixed this becomes
\beq
\rho\Delta x=\tfrac{1}{2}\lambda^2
\int_{-\infty}^\infty\!\dd x_1 \int_0^\infty\!\dd y_1
\int_{-\infty}^\infty\!\dd x_2 \int_0^\infty\!\dd y_2\,\,
\chi_{12}(\Delta x)\,\exp(-\lambda\pi y_{12}^2).
\label{eqnrDx2}
\eeq
The exponential factor in the integrand of (\ref{eqnrDx2}) 
has the interpretation
of the probability that the disk of radius $y_{12}$ around the point of
intersection $P_{12}$ contain no seeds.
For $i=1,2$ we now transform from the pair of variables $(x_i,y_i)$
to the new pair $(r_i,s_i)$ defined by
\bea
r_i \equiv f_i(0)  &=& \frac{y_i}{2}\Big( 1+\frac{x_i^2}{y_i^2} \Big),
\nonumber \\[2mm]
s_i \equiv f'_i(0) &=& -\frac{x_i}{y_i}, \qquad (i=1,2).
\label{defrisi}
\eea
The transformation has the Jacobian
\beq
\frac{\partial(r_i,s_i)}{\partial(x_i,y_i)} = \frac{(1+s_i^2)^2}{4r_i},
\qquad (i=1,2).
\eeq
In terms of these new variables of integration (\ref{eqnrDx2}) becomes
\bea
\rho\Delta x &=& 8\lambda^2
\int_0^\infty\!\dd s_1\int_{-\infty}^\infty\!\dd r_1
\int_0^\infty\!\dd s_2 \nonumber\\[2mm]
&& \times \int_{-\infty}^\infty\!\dd r_2
\frac{r_1r_2}{(1+s_1^2)^2(1+s_2^2)^2}\,\,
\chi_{12}(\Delta x)\,\exp(-\lambda\pi y_{12}^2).
\label{eqnrDx3}
\eea
On the interval $0\leq x\leq \Delta x$ the two parabolas $f_i(x)$
may be represented by the expansion
\beq
f_i(x) = r_i + s_i x + {\cal O}(\Delta x^2), \qquad (i=1,2).
\label{eqnfixlinear}
\eeq
The abscissa $x_{12}$ of their intersection point 
inside this interval, if there is one, is
the solution of $f_1(x_{12})=f_2(x_{12})$. 
Using (\ref{eqnfixlinear}) we find
\beq
x_{12}=\frac{r_2-r_1}{s_2-s_1} +{\cal O}(\Delta x^2).
\label{solnx12}
\eeq
The condition $0\leq x_{12} \leq \Delta x$ imposed by $\chi_{12}$ in 
(\ref{eqnrDx3}) can be rewritten as
\beq
r_1 \leq r_2 \leq  y_1+(s_1-s_2)\Delta x, \qquad (s_1>s_2).
\label{intcondy1y2}
\eeq
Because of symmetry we may restrict the $s_2$ integration in (\ref{eqnrDx3}) 
to $s_2<s_1$ if we compensate by an extra factor $2$. Doing so, interchanging
then the $r_2$ and $s_2$ integrals, using condition 
(\ref{intcondy1y2}), and still observing that $y_{12}=r_2+{\cal O}(\Delta x)$
 we obtain from (\ref{eqnrDx3})
\bea
\rho\Delta x &=& 16\lambda^2
\int_0^\infty\!\dd r_1\int_{-\infty}^\infty\!\dd s_1
\int_{-\infty}^{s_1}\!\dd s_2 \nonumber \\[2mm]
&& \times \int_{r_1}^{r_1+(s_1-s_2)\Delta x}\!\dd r_2
\frac{r_1r_2}{(1+s_1^2)^2(1+s_2^2)^2} \,\exp(-\lambda\pi r_1^2),
\label{eqnrDx4}
\eea
valid up to corrections of order $\Delta x^2$.
Dividing by $\Delta x$ and performing
the $r_2$ integral in the limit of $\Delta x\to 0$
we find
\beq
\rho=16\lambda^2 I_r(I_1-I_2)
\label{eqnrho}
\eeq
where
\bea
I_r &=&\tfrac{1}{2}\int_{-\infty}^\infty\!\dd r_1\,r_1^2\,
\exp(-\lambda\pi r_1^2) \nonumber \\[2mm]
&=& (4\pi\lambda^{\frac{3}{2}})^{-1}
\label{eqnIr}
\eea
and
\beq
I_i=\int_{-\infty}^\infty\!\dd s_1\int_{-\infty}^{s_1}\!\dd s_2\,\,
\frac{s_i}{(1+s_1^2)^2(1+s_2^2)^2}\,,
\qquad (i=1,2).
\label{eqnIi}
\eeq
In the expression for $I_1$ we interchange the $s_1$ and $s_2$ integrals,
whereas in the one for $I_2$ we interchange the names $s_1$ and $s_2$.
The result is that
\bea
I_1-I_2 &=& \int_{-\infty}^\infty \frac{\dd s_2}{(1+s_2^2)^2}
\,\Bigg( \int_{s_2}^\infty - \int_{-\infty}^{s_2} \Bigg)\,
\!\frac{s_1\,\dd s_1}{(1+s_1^2)^2} \nonumber\\[2mm]
&=& \int_{-\infty}^\infty \frac{\dd s_2}{(1+s_2^2)^2}
\,\Bigg[ -\frac{1}{2(1+s_1^2)} 
\Big( \Big|_{s_2}^\infty  - \Big|_{-\infty}^{s_2} \Big) \Bigg]
\nonumber\\[2mm]
&=& \int_{-\infty}^\infty \frac{\dd s_2}{(1+s_2^2)^3} \nonumber\\[2mm]
&=& \frac{3\pi}{8}.
\label{eqnIs}
\eea
Upon combining (\ref{eqnrho}), (\ref{eqnIr}), and (\ref{eqnIs}) we find
\bea
\rho &=& 16\lambda^2\times(4\pi\lambda^{\frac{3}{2}})^{-1}
\times(3\pi/8) \nonumber \\[2mm]
&=& \tfrac{3}{2}\lambda^{\frac{1}{2}},
\label{eqnrhofinal}
\eea
which is what we had to prove.\,\,$\square$



\begin{thebibliography}{10}
\bibitem{footnote1} 
Figure \ref{Fig2} is only schematic in that a faithful
  representation of all proportions would lead to an about twice
  higher density of the
  second and further neighbors to the right of $\Gamma_1$; as it is,
the picture better brings out the emergence
of the piecewise parabolic segments of $\Gamma_1$.
\bibitem{footnote2}
This is in agreement with the inequality $m_n>4$ for all $n$,
derived by Le Ca\"er and Delannay \cite{LeCaerDelannay93}.
\bibitem{footnote4}
More generally one might take a box $[-L,L]\times[0,M_L]$
where it suffices that $\lim_{L\to\infty}M_L=\infty$.
\end{thebibliography}
\end{document}